\documentclass[letterpaper,twocolumn,fleqn]{article} 

\usepackage{ist}
\usepackage{multirow}
\usepackage{graphicx}
\usepackage{xcolor}
\usepackage[caption=false]{subfig}
\usepackage{amsmath}
\usepackage{amsfonts}
\newcommand{\norm}[1]{\left\lVert#1\right\rVert}

\makeatletter
\renewcommand\p@subfigure{\thefigure~(}
\makeatother

\title{Image data hiding with multi-scale autoencoder network}

\author{Chen-Hsiu Huang and Ja-Ling Wu; Department of Computer Science and Information Engineering, Communication and Multimedia Laboratory, National Taiwan University, Taipei 106, Taiwan}

\date{} % date has an empty field.

\begin{document} 

\maketitle 

\thispagestyle{empty} % prevents the first page to be numbered

%%%%%%%%%%%%%%%%%%%%%%%%%%%%%%%%%%
% Abstract
%%%%%%%%%%%%%%%%%%%%%%%%%%%%%%%%%%

\begin{abstract}
Image steganography is the process of hiding information which can be text, image, or video inside a cover image. The advantage of steganography over cryptography is that the intended secret message does not attract attention and is thus more suitable for secret communication in a highly-surveillant environment such as civil disobedience movements. Internet memes in social media and messaging apps have become a popular culture worldwide, so this folk custom is a good application scenario for image steganography. We try to explore and adopt the steganography techniques on the Internet memes in this work. We implement and improve the HiDDeN model \cite{zhu2018hidden} by changing the Conv-BN-ReLU blocks convolution layer with a multiscale autoencoder network so that the neural network learns to embed message bits in higher-level feature space. Compared to methods that convolve feature filters on the row-pixel domain, our proposed MS-Hidden network learns to hide secrets in both low-level and high-level image features. As a result, the proposed model significantly reduces the bit-error rate to empirically 0\% and the required network parameters are much less than the HiDDeN model. 
\end{abstract}

%%%%%%%%%%%%%%%%%%%%%%%%%%%%%%%%%%%%
% Overall Document Guidelines: Head
%%%%%%%%%%%%%%%%%%%%%%%%%%%%%%%%%%%%
\section{Introduction} \label{sec:intro}

When two parties want to communicate with each other and preserve their privacy securely, the most commonly used strategy is data encryption. The data is converted into ciphertext using cryptography algorithms. The original message is not readable after encryption, but the ciphertext is visible to human eyes, leading to suspicion and further scrutiny. The advantage of steganography over cryptography is that the intended secret message does not attract attention. The visible encrypted messages, no matter how unbreakable they are, arouse interest and may in themselves be criminal in some countries \cite{map2021encryption}. Even worse, the two parties may be captured by an untrusted leagal authority in a civil disobedience movement \cite{wiki:Civil_disobedience} and being tortured with Rubber hose cryptanalysis \cite{wiki:rubberhosecrypt}. 

\subsection{Messsaging Apps}

\begin{figure}[!ht]
\centering
\subfloat[]{{\includegraphics[width=0.9\columnwidth]{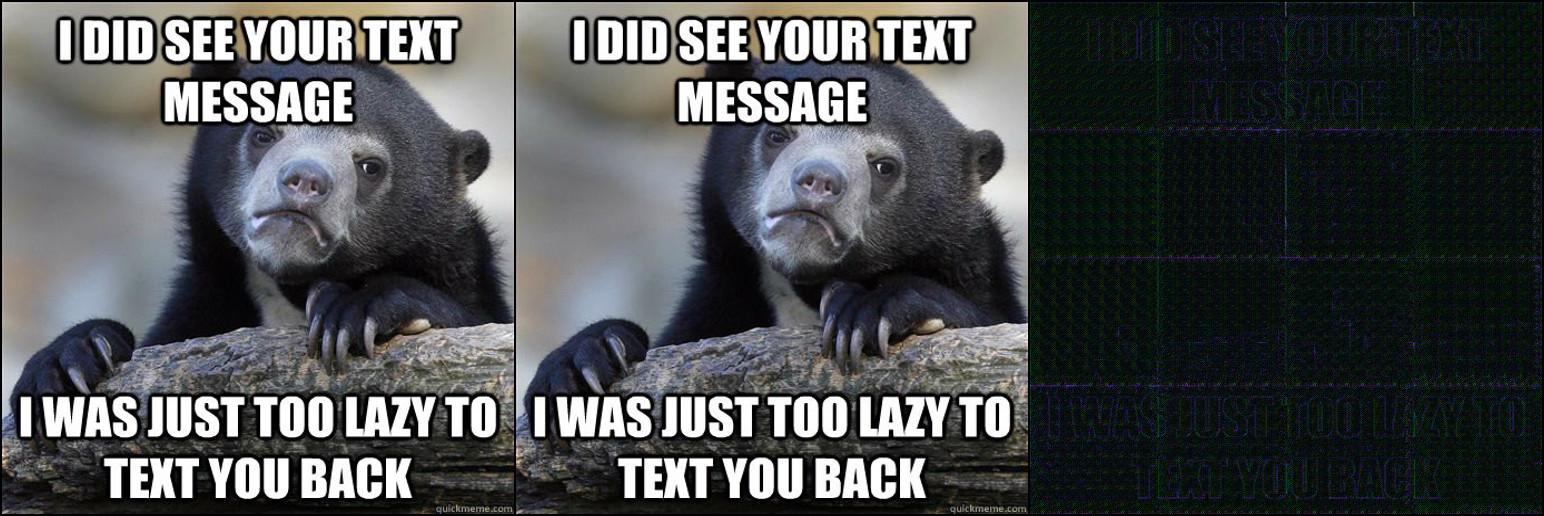} } }\\
\subfloat[]{{\includegraphics[width=0.9\columnwidth]{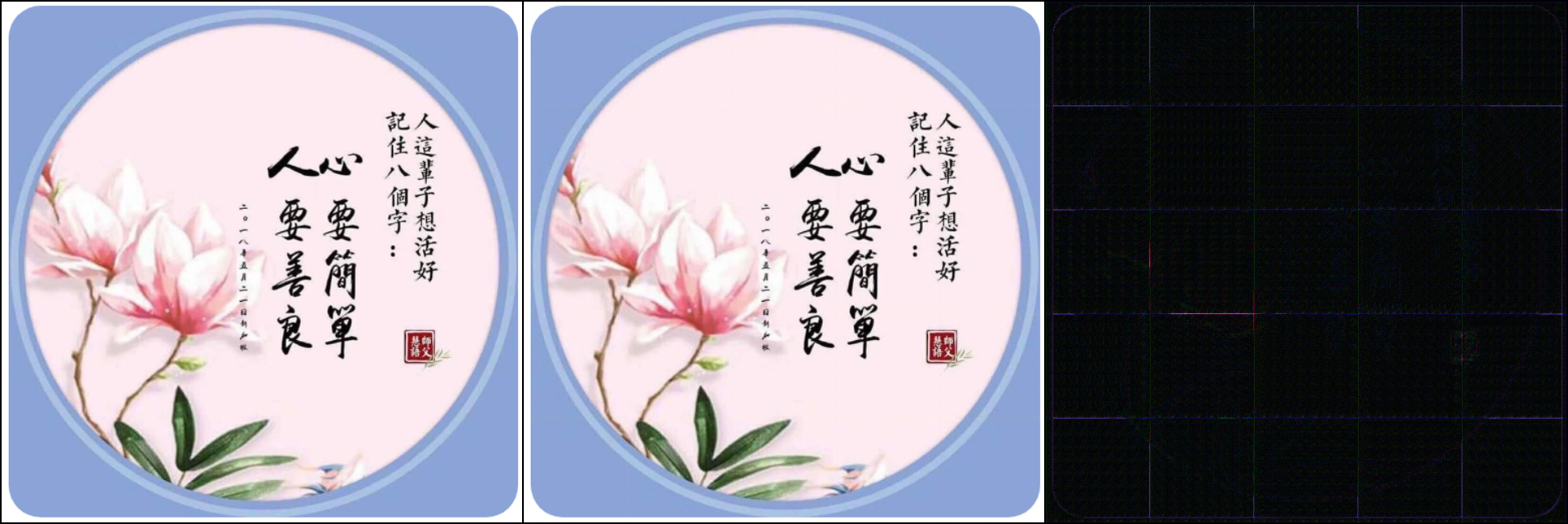} } \label{fig:cbai2-comp} }
\caption{Internet memes examples that shared among messaging apps. We embed the first 127 and 200 bytes of this paper's abstract text in images (a) and (b), respectively. The embedded text can be 100\% correctly extracted from the stego image with the proposed multi-scale autoencoder embedding network. From left to right, we show cover image ($C$), stego image ($C'$), and magnified difference as $|C-C'|\times 15$. }
\label{fig:changbei-ex}
\end{figure}

In the Mobile computing and Mobile Internet age, people with a highly cost-effective mobile phone can easily communicate with friends and families using instant messaging apps (such as Facebook Messenger and LINE). The modern messaging apps connect users to social groups and enable users to send messages instantly in all kinds of multimedia formats. The richness of images or photos significantly improves the user experience during communication, and thus, sending interesting photos becomes a dominant user behavior in messaging activities. Figure \ref{fig:changbei-ex} shows two examples of Internet memes shared among western and Asian societies. This kind of social phenomenon has become a folk custom, and the spread of Internet memes in messaging apps tends to be ignored by the ill-intentioned authority, which is a perfect application scenario for steganography.

\subsection{Steganography in Secret Communication}

Image steganography or watermarking is the process of hiding secrets inside a cover image for communication or proof of ownership. Zhu et al. \cite{zhu2018hidden} proposed the first deep learning-based image data hiding technique, the HiDDeN model, to achieve steganography and watermarking with the same neural network architecture. Except for HiDDeN, various DL-based image data hiding approaches have been proposed for steganography \cite{baluja2017hiding} (DDH) \cite{zhang2020udh} (UDH) \cite{tancik2020stegastamp} (StegaStamp) and watermarking \cite{luo2020distortion} \cite{luo2021dvmark} (DVMark) purpose. Our attempt to improve the HiDDeN model echos Baluja's \cite{baluja2017hiding} finding that the network architecture affects the secret hiding process and robustness of the DL-based methods. We want to explore the possibility of learning and hiding secret messages in both low-level and higher-level features representation of the encoder-decoder bottleneck layer. We use our network's convolution and downsampling strategies to learn features that constitute wider image spatial regions and reduce the required model parameters for faster training and embedding.

\begin{figure}[!ht]
\centering
\subfloat[HiDDeN]{{\includegraphics[width=0.45\columnwidth]{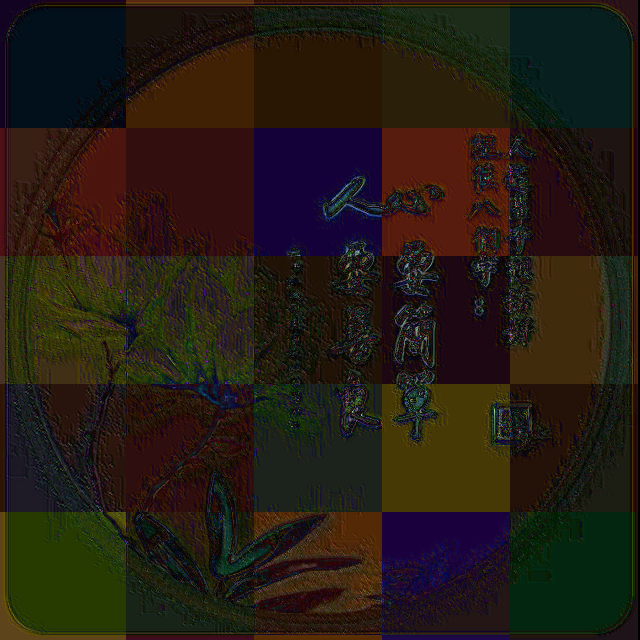} } \label{fig:hid-diff} } 
\subfloat[Proposed]{{\includegraphics[width=0.45\columnwidth]{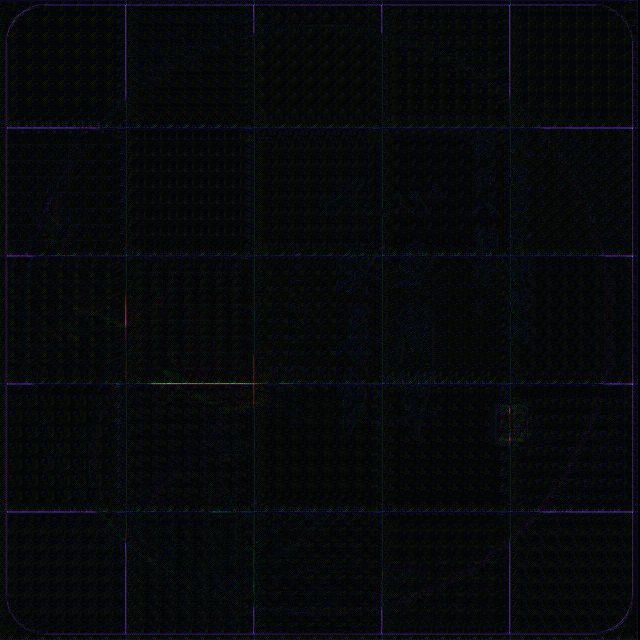} } \label{fig:msae-diff} }
\caption{The magnified embedding differences of the HiDDeN model and the proposed model on the testing image, shown in Figure \ref{fig:cbai2-comp}).}
\label{fig:hidden-low}
\end{figure}

The HiDDeN model uses a sequence of Conv-BN-ReLU blocks that only learn from low-level image features and uniformly hide data into raw pixel domain without considering higher-level features such as edges or regions. Furthermore, the model complexity of deep Conv-BN-ReLU layers limits the use of deep learning-based steganography on larger-sized input images; moreover, it requires block-based data hiding that may cause extra blocking effects. The two disadvantages can be demonstrated in Figure \ref{fig:hid-diff}), where we see apparent low-level pixel differences between neighboring blocks on smooth regions. Ideally, the embedding network with a similar transform should alter the monotonic continuous areas, like what we observe in Figure \ref{fig:msae-diff}). 

%%%%%%%%%%%%%%%%%%%%%%%%%%%%%%%%%%
% Graphics and Equations
%%%%%%%%%%%%%%%%%%%%%%%%%%%%%%%%%%
\section{Related Works}

The image steganography techniques have been developed with a long history. Many traditional methods are based on the LSB (least significant bit) technique or custom-designed algorithms, such as \cite{pevny2010using} (HUGO) \cite{holub2014universal} (S-UNIWARD), or change mid-frequency components in the frequency domain \cite{bi2007robust} (WOW).

Deep steganography methods such as DDH \cite{baluja2017hiding} and UDH \cite{zhang2020udh} defined a new task to hide a whole image in another with a deep neural network. Unlike traditional steganography that requires a perfect restore of secret messages, DDH and its variants minimize the distortion between the retrieved and the original secret images. Thus, the message is securely delivered because the authentic and recovered secret images are visually indistinguishable. Lu et al. \cite{lu2021large} recently advanced deep steganography with a higher capacity of up to three or more secret images.

Zhu et al. \cite{zhu2018hidden} proposed the HiDDeN model that embeds the raw bits and extracts the secret message with a low bit-error rate using a deep neural network. With the generative model adversarial trained against the noise layers, the HiDDeN model can achieve the purposes of steganography and digital watermark with the same network architecture. Perhaps inspired by HiDDeN, many researchers have proposed similar adversarial network methods for steganography \cite{tancik2020stegastamp} \cite{jing2021hinet} and watermarking \cite{luo2020distortion} \cite{luo2021dvmark}. 

\section{Proposed Methods}

\begin{figure}[!ht]
\centering
\includegraphics[width=\columnwidth]{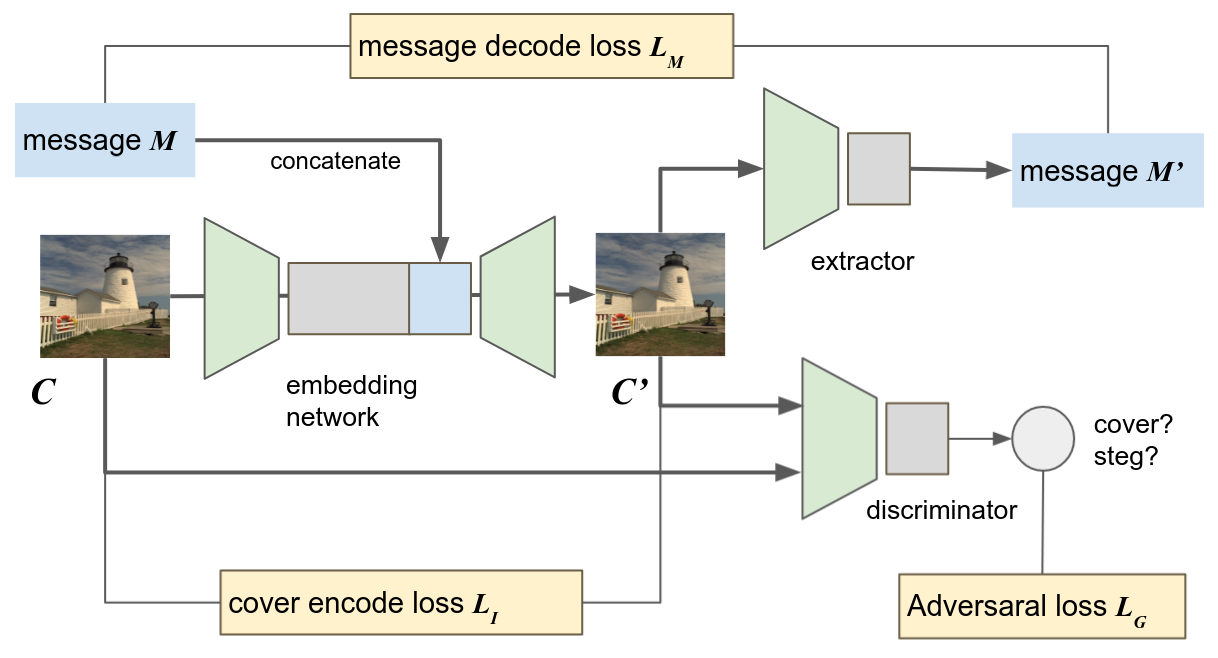}
\caption{The overall architecture of the proposed MS-Hidden. }
\label{fig:msae-all-arch}
\end{figure}

We use the publically available HiDDeN\footnote{https://github.com/ando-khachatryan/HiDDeN} implementation and modify\footnote{Our code is available at https://github.com/chenhsiu48/HiDDeN} the Conv-BN-ReLU blocks convolution layer with the proposed multi-scale autoencoder network. Figure \ref{fig:msae-all-arch} shows the overall proposed architecture. Compared with the HiDDeN model, we revise the embedding network and remove the noise layer because robustness is not considered in steganography.

The learning-based steganography methods use the convolution layers to transform the cover images to high-dimensional feature space for combining with the input secrects. Either message appending \cite{tancik2020stegastamp} or message concatenation \cite{zhu2018hidden} \cite{luo2020distortion} \cite{luo2021dvmark} has been widely adopted to integrate with the latent code representation and trained as residual toward the skipped connection of cover image. Most literatures use a sequence of Conv-BN-ReLU blocks \cite{baluja2017hiding} \cite{zhu2018hidden} \cite{zhang2020udh} or U-Net \cite{ronneberger2015u} liked network architecture \cite{luo2021dvmark} \cite{luo2020distortion} to learn this complex transformation.

\begin{figure*}[!ht]
\centering
\includegraphics[width=\textwidth]{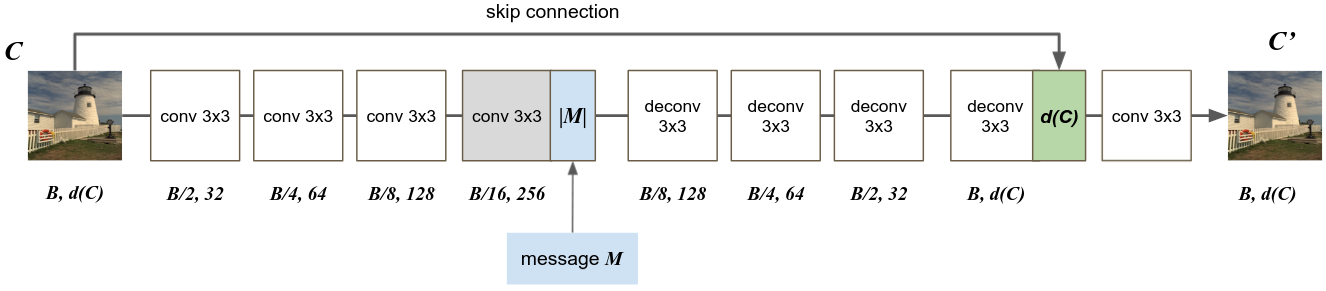}
\caption{The autoencoder architecture of our embedding network with $k=4$. Here $d(\cdot)$ retrieves the cardinality of the cover image components. } 
\label{fig:msae-enc-arch}
\end{figure*}

In our embedding network, we use the autoencoder network with $k$ convolution layers that downsamples the cover image $C$ into half and doubles the filter channels in each layer, as shown in Figure \ref{fig:msae-enc-arch} where $k=4$. The secret message $M$ with bits $|M|$ is duplicated and concatenated with the latent code representation. Then, the same deconvolution layers are applied on the latent code to restore back to the original resolution as residual to the cover image $C$. At least, a final convolution layer is used to blend the residual and the skipped image to the encoded stego image $C'$.

The extractor and discriminator share the same $k$ convolution layers as the encoder part of the embedding network, i.e., feature maps are spatially downsampled in half size and doubled in channels. The bottleneck layer of the extractor and discriminator then feeds to a linear layer of $|M|$ nodes and one node as final repression output, which is identical to the $|M|$ bits decoded message and classification result, respectively. All the convolution layers use a $3 \times 3$ kernel with stride 2.

\subsection{Loss Function}

To end-to-end learn the steganography model, we characterize the image distortion loss $L_I$ with $l_2$ distance between cover image $C$ and stego image $C'$ as $L_I(C,C')= \norm{C-C'}_2^2$. Similarly, we measure the message distortion loss $L_M$ with $l_2$ distance between the original and decoded message as $L_M(M,M')= \norm{M-M'}_2^2$. We adopt the adversal training with a discriminator $A$ that predicts $A(\hat{I}) \in [0,1]$, which is the probability of cover image, where $\hat{I} \in \{C,C'\}$. Then the adversarial loss $L_G$, the ability of the discriminator to detect stego image, is model as $L_G(C')=\log(1-A(C'))$. We use the PyTorch built-in Adam optimizer to train the embedding network, message extractor, and discriminator to minimize the loss over training images and random sampled message bits:

\begin{equation}
\mathbb{E}_{C,M}= \lambda_I \cdot L_I(C,C') + \lambda_M \cdot L_M(M,M')+ \lambda_G \cdot L_G(C') 
\end{equation}

where $\lambda_I=1.0$, $\lambda_M=1.5$ and $\lambda_G=0.001$ are hyperparameters. 

\subsection{Feature-Capacity Hypothesis}

From Figure \ref{fig:msae-enc-arch}, we can derive that the number of feature channels of the last convolution layer $f_c$ is $(2^k)^2$. Therefore, the number of channels of the bottleneck layer before deconvolution is $f_c + m_c$, where $m_c$ denotes the number of message channels, which equals $|M|$ in number. The spatial resolution of bottleneck layer is reduced from input image size $B \times B$ to feature maps of size $(B/2^k) \times (B/2^k)$. We define the feature-capacity ratio $\rho$ as:

\begin{equation}
\rho_{k,|M|} = \frac{m_c}{fc} = \frac{|M|}{(2^k)^2}
\end{equation}

It is intuitive to assume that the relative ratio between message channels $m_c$ and feature channels $f_c$ has impact on the overall network training efficiency and stability. Therefore, we have two hypotheses as follows.

\textbf{Hypothesis 1.} With the same convolution layers $k$, the more message bits $|M|$ to hide, the higher bit-error-rate we will obtain under the similar cover vs. stego image distortion. That is, for the same downsampling lays $k$, we have $E_{k,128} > E_{k,64}$.

\textbf{Hypothesis 2.} With the same feature-capacity ratio $\rho$ calculated from different convolution layers $k$ and message bits $|M|$, the converged network training result will have a similar bit-error-rate. That is, $\rho_{4,64} = 0.25 = \rho_{3,16} = \rho_{5,256}$.

In the HiDDeN model, the number of feature channels $f_c$ is fixed at 64, while message channels vary from 30 to 52. With the introduced feature-capacity ratio, we have a guideline to simplify our network architecture by reducing the autoencoder layers with lesser message bits.

\section{Experimental Results}

We randomly selected 20,000 images from the COCO dataset \cite{russakovsky2015imagenet} as the training set and 1,000 images for validation and testing set to train our model. We cross-validate our model on the BOSS \cite{bas2011break} dataset, ImageNet \cite{linmicrosoft}, and a custom-collected MEME dataset of 100 images to evaluate our model's generalization ability. We sampled 1,000 images from the BOSS and ImageNet datasets as our benchmark dataset.

We train our model with the PyTorch built-in Adam optimizer using a learning rate of 0.001 and batch size 30. We train our model for 150 epochs until the bit-error rate on the testing set is converged to 0\% and we pick the best model with the lowest encoding distortion. We compare our model with two DL-based steganography methods, HiDDeN and UDH. We follow the same training settings from the HiDDeN paper and train for 300 epochs. We use the official implementation\footnote{https://github.com/ChaoningZhang/Universal-Deep-Hiding} of UDH and download the pre-trained model on their website.

\subsection{Capacity and Secrecy}

\begin{figure}[!ht]
\centering
\subfloat[ILSVRC2012\_val\_00000029]{{\includegraphics[width=0.95\columnwidth]{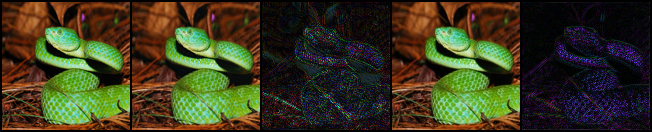} } \label{fig:im29}} \\
\subfloat[ILSVRC2012\_val\_00000562]{{\includegraphics[width=0.95\columnwidth]{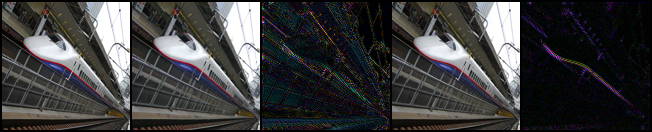} } \label{fig:im562}} \\
\subfloat[ILSVRC2012\_val\_00003310]{{\includegraphics[width=0.95\columnwidth]{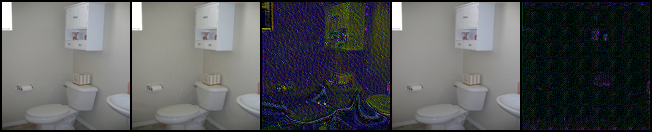} } \label{fig:im3310} }
\caption{Qualitative evaluation for the proposed MS-Hidden on selected ImageNet images. From left to right : cover image $C$, stego image $C_H'$ and residual image from HiDDeN, stego image $C'$ and residual image from proprosed method.} 
\label{fig:hhcomp}
\end{figure}

\textbf{Qualitative evaluation:} We resize the input cover image as $128 \times 128$ and generate the stego image for comparison. From the stego residual images of HiDDeN in Figure \ref{fig:hhcomp}, we see that the secrets are uniformly embedded in the raw-pixel domain, including smooth textures like the white wall in Figure \ref{fig:im3310}). At the same time, our MS-Hidden learns to hide secrets in high-level features primarily, e.g., the snakeskin texture in Figure \ref{fig:im29}) and the red stripe on the railroad car in Figure \ref{fig:im562}).

In our multi-scale autoencoder network, the spatial downsampling of feature maps in each convolution layer helps learn features on a broader reception field, resulting in higher-level latent representation. Because the secret bits are blended with multi-scale latent codes, our model learns to hide data in edges, textures (Figure \ref{fig:im29})), or regions (Figure \ref{fig:im562})) depending on the image content. For a less complex image with few details shown in Figure \ref{fig:im3310}), the embedding network chooses to hide messages uniformly in a periodic pattern.

\textbf{Quantitative evaluation:} We divide the cover image into $128 \times 128$ blocks and embed message bits to each block to maximize the total message volume for secret communication. We embed 64 bits message per $128 \times 128$ RGB image block for our MS-Hidden model, which yields 0.0013 BPP (bits per pixel). On the HiDDeN model, we tried to embed 64 bits messages per block but could not obtain the similar bit error rate mentioned in the original paper. Therefore, we choose to embed 32 bits per block (about 0.0007 BPP) to achieve a meaningful bit error rate close to the reported $10^{-5}$. 

Contrary to the HiDDeN-liked model that hides raw bits of the message into an image and relies on bit-error rate for communication, the UDH model hides one secret embodiment into another cover image in the same resolution, which is about 24 BPP capacity. However, since most of the secret image pixels extracted from the UDH stego image are altered, the bit-error rate is less meaningful. In UDH, the message is delivered because the hidden and the recovered secret images are visually indistinguishable. We modify the UDH to take the MSB (most significant bit) of the secret image as a binary message and measure the bit-error rate of recovered MSB, which is about 3 BPP in capacity.

We use three metrics to measure the quality of cover/stego image pairs. They are Peak Signal-to-Noise Ratio (PSNR), Structural Similarity Index (SSIM) \cite{wang2004image}, and Mean Absolute Error (MAE). Table \ref{tab:steg-perf} shows that our proposed MS-Hidden has an overall lower bit-error rate compared to HiDDeN and UDH. Conceptually the DL-based steganography methods can not achieve a 0\% bit error rate and require error-correcting code to enhance the recovering results. To our surprise, our model achieves a 0\% bit error rate during training and validation processes on the testing sets. We repeat testing the benchmark datasets on MS-Hidden 100 times and report the average bit error rate on Table \ref{tab:steg-perf}. Empirically, we demonstrate that the MS-Hidden model achieves close to 0\% bit error rate with the multi-scale autoencoder network. 

\begin{table}[ht]
\caption{The message bit-error rate and cover vs. stego image distortion comparison. We empirically demonstrate that our proposed method achieves close to 0\% bit-error rate with less image distortion.}
\label{tab:steg-perf}
\resizebox{\columnwidth}{!}{%
\begin{tabular}{|l|l|r|r|r|r|}
\hline
Dataset & Method & \multicolumn{1}{l|}{Bit-error} & \multicolumn{1}{l|}{PSNR} & \multicolumn{1}{l|}{SSIM} & \multicolumn{1}{l|}{MAE} \\ \hline
\multirow{3}{*}{BOSS} & HiDDeN & 0.000098 & 43.94 & 0.9915 & 1.20 \\ \cline{2-6} 
 & UDH & 0.007704 & 39.58 & 0.9348 & 2.27 \\ \cline{2-6} 
 & Proposed & \textbf{0.000000} & \textbf{48.23} & \textbf{0.9918} & \textbf{0.65} \\ \hline
\multirow{3}{*}{MEME} & HiDDeN & 0.009379 & 43.76 & \textbf{0.9941} & 1.21 \\ \cline{2-6} 
 & UDH & 0.006596 & 39.66 & 0.9343 & 2.15 \\ \cline{2-6} 
 & Proposed & \textbf{0.000000} & \textbf{46.08} & 0.9876 & \textbf{0.81} \\ \hline
\multirow{3}{*}{ImageNet} & HiDDeN & 0.001852 & 41.85 & \textbf{0.9888} & 1.53 \\ \cline{2-6} 
 & UDH & 0.009289 & 39.58 & 0.9463 & 2.17 \\ \cline{2-6} 
 & Proposed & \textbf{0.000000} & \textbf{44.44} & 0.9880 & \textbf{1.02} \\ \hline
\end{tabular}%
}
\end{table}

\subsection{Steganographic analysis}

Our steganalysis measured the possibility of distinguishing a stego image from a cover image using publicly available steganalysis tools, including traditional statistical methods \cite{boehm2014stegexpose} and new DL-based approaches \cite{lerch2016unsupervised} \cite{boroumand2018deep}. We use the steganalysis tool, StegExpose \cite{boehm2014stegexpose} to measure our model's anti-steganalysis ability. We randomly select half of our 1,000 ImageNet's testing set to generate the stego images, and the other half use the cover images to benchmark with StegExpose.

\begin{figure}[!ht]
\centering
\includegraphics[width=0.9\columnwidth]{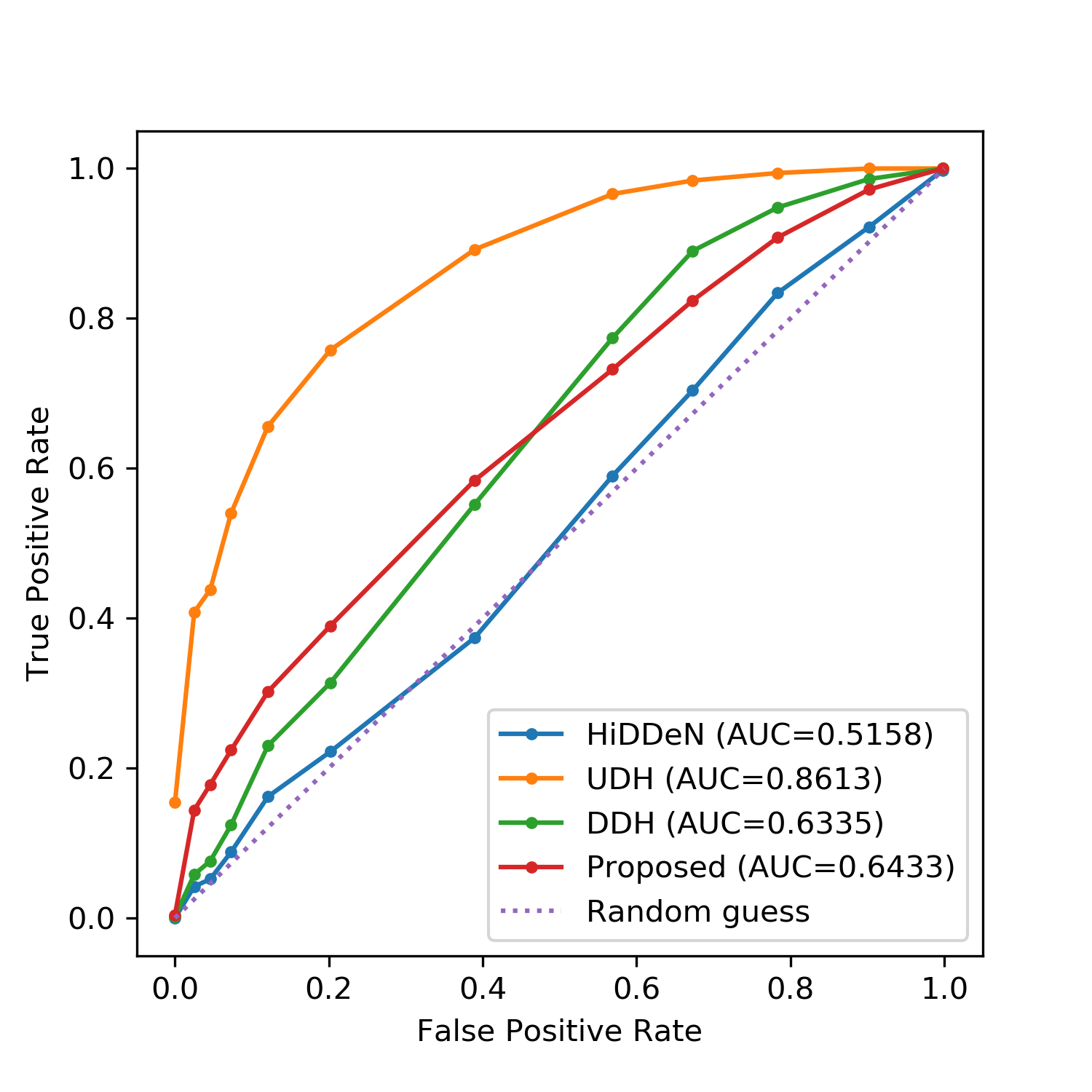}
\caption{The ROC curve comparison produced by StegExpose.}
\label{fig:msae-roc}
\end{figure}

To draw the ROC (receiver operating characteristic) curve, we vary the detection thresholds as input to StegExpose and come out the ROC curve. Figure \ref{fig:msae-roc} shows the ROC curve of the HiDDeN model, UDH, and our proposed method. We calculate the AUC (area under the curve) of each ROC curve and see that the HiDDeN model has the best AUC as 0.5158, indicating that the stego image detection accuracy is quite close to random guess. The AUC of our proposed model is inferior to HiDDeN; we think that may arise due to the causal relationship between high-level features and the pixel modification that makes StegExpose easier to detect. As for UDH, due to the trade-off between capacity and secrecy, this deep steganography method can be easily detected by the StegExpose algorithm.

\subsection{Feature-Capacity Hypothesis} 

The above experiments show that we can effectively hide 64 bits of messages into a $128 \times 128$ image with an embedding network of $k=4$, where the feature-capacity ratio $\rho_{4,64}=64/256=0.25$. The same $\rho_{3,16}=16/(2^3)^2=0.25$ can be calculated if we hide 16 bits message into a $32 \times 32$ image with $k=3$. We empirically validate the Hypothesis 2 in the second row of Table \ref{tab:fcratio}, where the BPP of $\rho_{3,16}$ is four times than $\rho_{4,64}$, but the averaged bit-error remains close to 0\%.

As we increase the message from 64 bits to 128 bits onto $128 \times 128$ image block with $k=4$ (the third row of Table \ref{tab:fcratio}), we have: 

\begin{equation}
\rho_{4,128}=\frac{128}{256}=0.5 > \rho_{4,64}=\frac{64}{256}=0.25
\end{equation}

As a result, the bit-error rate $E_{4,128} = 0.02 > E_{4,64} = 0.0$, which validates Hypothesis 1. Furthermore, if we increase the convolution layers of the multi-scale autoencoder with $k=5$, the bit-error rate $E_{5,128}=0.0016<E_{4,128}=0.02$, due to $\rho_{5,128}=0.125<\rho_{4,128}=0.5$. The feature-capacity hypothesis hints us that we can add more convolution layers of the autoencoder to increase the embedding capacity. However, we will encounter network training difficulty in reality and thus not always applicable.

\begin{table}[!ht]
\caption{Embedding comparison of different feature-capacity ratio settings on ImageNet testing set.}
\label{tab:fcratio}
\resizebox{\columnwidth}{!}{%
\begin{tabular}{|l|r|r|r|r|}
\hline
Method & \multicolumn{1}{l|}{BPP} & \multicolumn{1}{l|}{Bit-error} & \multicolumn{1}{l|}{PSNR} & \multicolumn{1}{l|}{MAE} \\ \hline
$B$=128, $k$=4, $|M|$=64 & 0.0013 & 0.000000 & 44.44 & 1.02 \\ \hline
$B$=32, $k$=3, $|M|$=16 & 0.0052 & 0.000000 & 35.43 & 3.05 \\ \hline
$B$=128, $k$=4, $|M|$=128 & 0.0026 & 0.020392 & 38.12 & 2.59 \\ \hline
$B$=128, $k$=5, $|M|$=128 & 0.0026 & 0.001694 & 36.07 & 3.09 \\ \hline
\end{tabular}%
}
\end{table}

\subsection{Model Complexity}

To the best of our knowledge, there is no steganography method in the literature to study the model complexity vs. data hiding secrecy. Beyond traditional capacity vs. secrecy evaluation, we think it's essential to consider the computational cost of DL-based steganography methods, given today's high quality and substantial volume images. We calculate the model complexity of each comparing DL-based processes and measure the embedding performance on an Intel Core i7-9700K CPU workstation with an Nvidia GeForce RTX 2080 Ti GPU.

Table \ref{tab:embed-perf} shows the message embedding performance and the corresponding model complexity of the proposed MS-Hidden network compared to the HiDDeN model. The embedding time cost is measured in seconds and averaged per image across all testing datasets. With the multi-scale convolution layers that downsample each unique feature into compact latent representation, our proposed model reduces the required FLOPS in magnitude 17.8 times more efficiently. As a result, the speed to embed message per image is accelerated 3.4 times than the HiDDeN model.

\begin{table}[ht]
\caption{The message embedding performance and model complexity comparison.}
\centering
\label{tab:embed-perf}
\resizebox{0.85\columnwidth}{!}{%
\begin{tabular}{|l|r|r|r|}
\hline
Method & \multicolumn{1}{l|}{Time (sec.)} & \multicolumn{1}{l|}{Params (M)} & \multicolumn{1}{l|}{FLOPS (G)} \\ \hline
HiDDeN & 0.1149 & 0.4143 & 6.7718 \\ \hline
Proposed & 0.0333 & 1.2553 & 0.3834 \\ \hline
\end{tabular}%
}
\end{table}

\section{Conclusion}

This work focuses on embedding raw bits of messages in the popular Internet memes shared in social media and messaging apps. Applying steganography techniques on Internet memes for secret communication tends to be ignored by the ill-intentioned authority, which is a perfect application scenario for steganography. We explore the possibility of learning and hiding secret messages in a multi-scale feature representation of the cover image. We implement and improve the HiDDeN model by changing the Conv-BN-ReLU blocks convolution layer with a multi-scale autoencoder network. Our autoencoder network is parameterized by the number of layers $k$ and message bit-length $|M|$; therefore, the feature-capacity ratio can be defined as a guideline to adjust the convolution layers with corresponding message length.

Compared with methods that convolve feature filters on the row-pixel domain, our proposed MS-Hidden network learns to hide secrets in multi-scale image features and significantly reduces the bit-error rate to empirically 0\%. Our stego image distortion is superior to that of the compared methods in both qualitative and quantitative evaluations. With the multi-scale convolution layers that downsample each specific feature into compact latent representation, our proposed model reduces the required FLOPS in an order of magnitude and significantly accelerates the message embedding performance.

%%%%%%%%%%%%%%%%%%%%%%%%%%%%%%%%%%
% Reference Preparation
%%%%%%%%%%%%%%%%%%%%%%%%%%%%%%%%%%

%\section{Reference Preparation}
%Use the standard LaTeX \emph{cite} command for references in the text. You can then use the standard bibliography command to generate the list of references. Add the command \emph{small} before the bibliography to give it the right font size.  Reference \cite{bib1} style should be used for books, Reference \cite{bib2} style should be used for Journals, and Reference \cite{bib3} style should be used for Proceedings.

\section{Acknowledgments} 

The authors would like to thank the MOST of Taiwan and CITI SINICA for supporting this research under the grant numbers MOST 108-2218-e-002-055, MOST 108-2221-E-002-103-my3, and Sinica 3012-C3447.


%%%%%%%%%%%%%%%%%%%%%%%%%%%%%%%%%%
% Bibliography
%%%%%%%%%%%%%%%%%%%%%%%%%%%%%%%%%%

\bibliographystyle{plain} 
\small
\bibliography{refs}

%\small
%\begin{thebibliography}{9}
%\bibitem{bib1}John Doe, Recent Progress in Digital Halftoning II, IS\&T, Springfield, VA, 1999, pg. 173.
%\bibitem{bib2}John Doe, Digital Imaging, J. Imaging. Sci. and Technol., 42, 112 (1998).
%\bibitem{bib3}John Doe, An Inexpensive Micro-Goniophotometry You Can Build, Proc. PICS, pg. 179. (1998).
%\end{thebibliography}


\begin{biography}
Chen-Hsiu Huang received the B.S. and M.S. degrees in computer science from the Department of Computer Science and Information Engineering (CSIE), National Taiwan University (NTU), in 2002 and 2004. Since that, he has worked in the software industry for 16 years, joined various companies with domains including media entertainment, computer security, e-commerce, and video streaming. He is now pursuing a Ph.D. degree with the Department of CSIE, NTU.

Ja-Ling Wu has been a professor with the Department of Computer Science and Information Engineering, National Taiwan University (NTU) since 1996. From 2004 to 2007, he was the Head of the Graduate Institute of Networking and Multimedia, NTU. He was selected to be one of the life-time Distinguished Professors of NTU in 2006. Professor Wu has been a Fellow and a Life-fellow of IEEE since 2008 and 2022.
\end{biography}

\end{document}